\documentclass[%
 reprint,
showpacs,preprintnumbers,
 amsmath,amssymb,
 aps,
prl,
]{revtex4-1}

\usepackage{graphicx}
\usepackage{dcolumn}
\usepackage{bm}
\usepackage{color}
\usepackage{ulem}

\begin{document}

\preprint{APS/123-QED}

\title{
Multiple-$Q$ Instability by ($d-2$)-dimensional Connections of Fermi Surfaces 
}

\author{Satoru Hayami and Yukitoshi Motome}
\affiliation{%
 Department of Applied Physics, University of Tokyo, Tokyo 113-8656, Japan 
}%

\begin{abstract}
We propose a general mechanism of multiple-$Q$ ordering with noncollinear and noncoplanar spin textures in itinerant magnets. 
By analyzing the fourth-order perturbation with respect to the spin-charge coupling, we find that the instability toward multiple-$Q$ ordering is caused by ($d-2$)-dimensional connections of the Fermi surfaces in the extended Brillouin zone by the multiple-$Q$ wave vectors: zero(one)-dimensional point (line) connections in the two(three)-dimensional systems. 
The instabilities are obtained as the ``fixed points" in the two-parameter flow diagram, irrespective of lattice structures. 
The hidden Fermi surface instability provides a universal origin of noncollinear and noncoplanar instabilities common to frustrated and unfrustrated lattice systems. 
\end{abstract}
                            
\maketitle

The nesting property of Fermi surface (FS) plays an important role in understanding of the instabilities occurring in itinerant electron systems~\cite{peierls1955quantum,RevModPhys.60.1129,RevModPhys.66.1}. 
For instance, an instability toward electronic ordering, such as charge density wave and spin density wave, occurs predominantly at the nesting wave vector. 
A lattice distortion is also caused by the nesting, called the Peierls instability. 
In the case of the perfect nesting, in which all the Fermi vectors are connected with others by a single nesting vector, the system is unstable against an infinitesimal perturbation; e.g., a N\'eel order with the ($\pi,\pi$) wave vector is immediately induced by introducing Coulomb interaction in the tight-binding model on a square lattice at half filling where the square-shaped FS is perfectly nested. 

The perfect nesting also leads to peculiar noncollinear and noncoplanar ordering. 
This happens when the FS is nested by more than a single wave vector. 
An example was found in a two-dimensional (2D) triangular lattice system at 3/4 filling~\cite{Martin:PhysRevLett.101.156402}. 
In this case, the FS is perfectly nested by three wave vectors. 
This special nesting leads to an instability toward triple-$Q$ magnetic ordering characterized by the three wave vectors, which has a peculiar noncoplanar spin texture. 
Another example was found in a three-dimensional (3D) pyrochlore lattice system at 1/4 filling~\cite{Chern:PhysRevLett.105.226403}. 
The FS consists of lines on the Brillouin zone boundaries, which are perfectly nested by three wave vectors, and a complicated noncoplanar spin order was predicted also in this case. 
In the checkerboard lattice system at 1/4 filling, a coplanar but noncollinear double-$Q$ order is induced by the perfect nesting~\cite{Venderbos:PhysRevLett.109.166405}. 
These noncollinear and noncoplanar magnetic orders have attracted interests as they 
induce unusual transport properties, such as unconventional Hall effect~\cite{Loss_PhysRevB.45.13544,Ye_PhysRevLett.83.3737, Ohgushi:PhysRevB.62.R6065,taguchi2001spin} 
and spin Hall effect~\cite{TaguchiPhysRevB.79.054423}.

On the other hand, similar noncollinear and noncoplanar orders are found even in the absence of perfect nesting. 
Also they appear on both geometrically-frustrated and unfrustrated lattices. 
For instance, a noncollinear double-$Q$ order, similar to the checkerboard case, was found in a square lattice system at 1/4 filling~\cite{Yunoki:PhysRevB.62.13816}. 
It was also suggested that a triple-$Q$ noncoplanar order is realized on a face-centered-cubic (fcc) lattice~\cite{Yunoki:PhysRevB.62.13816}. 
Meanwhile, another triple-$Q$ noncoplanar state is stabilized in a cubic lattice system at 1/4 filling~\cite{Alonso:PhysRevB.64.054408,Hayami:PhysRevB.89.085124}. 
In the triangular lattice case also, the same noncoplanar triple-$Q$ state as that by the perfect nesting at 3/4 filling is stabilized at 1/4 filling~\cite{akagi2010spin,Kumar:PhysRevLett.105.216405,Kato:PhysRevLett.105.266405}. 
In all these cases, the FSs are rather small, far from the perfect nesting.

The fact that noncollinear and noncoplanar instabilities are ubiquitously observed in various lattice structures strongly suggests a general mechanism beyond the perfect nesting of FS. 
The mechanism is distinct from the spin-orbit coupling (Dzyaloshinskii-Moriya interaction), which is relevant to noncollinear orders in multiferroics~\cite{Fiebig:0022-3727-38-8-R01,Katsura:PhysRevLett.95.057205,Sergienko_PhysRevB.73.094434,Jia:PhysRevB.76.144424,khomskii2009classifying,PhysRevLett.96.067601} and triple-$Q$ noncoplanar orders in skyrmion crystals~\cite{muhlbauer2009skyrmion,yu2010real}. 
In the previous study by one of the authors and his collaborators~\cite{akagi2012hidden}, the importance of FS connections was pointed out for the frustrated triangular lattice case, but it remains elusive whether it is generically applicable to other cases, especially on unfrustrated lattices.  
For further exploration of the physics associated with noncollinear and noncoplanar ordering, 
it is highly desirable to establish unified understanding of the stabilization mechanism of the multiple-$Q$ states.

\begin{figure}[t!]
\begin{center}
\includegraphics[width=1.0 \hsize]{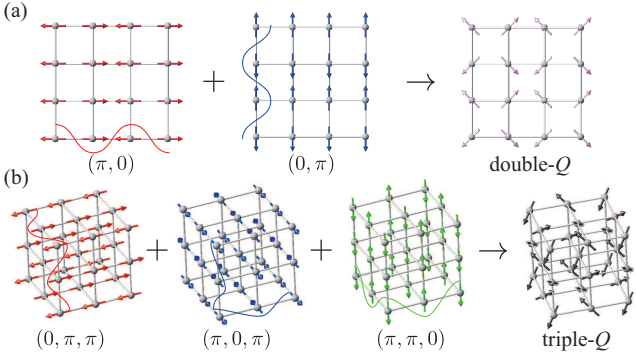} 
\caption{
\label{Fig:ponti}
(Color online) 
Schematic pictures of (a) a double-$Q$ state on a 2D square lattice and (b) a triple-$Q$ state on a 3D cubic lattice. 
The arrows denote the directions of local magnetic moments. 
}
\end{center}
\end{figure}

In this paper, we elucidate the general mechanism of the multiple-$Q$ instabilities in itinerant electrons. 
For noninteracting electrons coupled with localized spins, we derive the general form of the free energy by the fourth-order perturbation with respect to the spin-charge coupling. 
We find that the result is summarized by only two parameters, irrespective of the lattice structures as well as the system dimensions. 
We show that the multiple-$Q$ instabilities occur at particular electron filling, where $(d-2)$-dimensional portions of the FS are connected by the multiple-$Q$ wave vectors ($d$ is the system dimension). 
This is distinct from the perfect nesting that corresponds to a $(d-1)$-dimensional connection. 
We also provide a versatile graphical representation for identifying the instabilities in the form of the flow diagram of the two relevant parameters.

Let us begin with defining the multiple-$Q$ orders that we consider in this study. 
The order parameters for double- and triple-$Q$ states are written by 
\begin{align}
\label{eq:multiple-Q}
\langle \bm{S}_{i} \rangle \propto \left\{
\begin{array}{l} 
(\cos \bm{Q}_1 \cdot \bm{r}_i, \cos \bm{Q}_2 \cdot \bm{r}_i, 0), \\ 
(\cos \bm{Q}_1 \cdot \bm{r}_i, \cos \bm{Q}_2 \cdot \bm{r}_i, \cos \bm{Q}_3 \cdot \bm{r}_i), 
\end{array}
\right.
\end{align}
respectively. 
Here, $\bm{Q}_1$, $\bm{Q}_2$, and $\bm{Q}_3$ stand for the wave vectors characterizing the multiple-$Q$ states; $\bm{r}_i$ is the position vector of the site $i$. 
Equation~(\ref{eq:multiple-Q}) defines noncollinear and noncoplanar multiple-$Q$ magnetic orders represented by superpositions of single-$Q$ states for different spin components. 
We here restrict ourselves to the superposition with an equal weight.
For instance, a double-$Q$ state on a 2D square lattice is given by the superposition of ($\cos \pi r_i^x,0,0$) and ($0, \cos \pi r_i^y,0$), as shown in Fig.~\ref{Fig:ponti}(a), which is described by taking $\bm{Q}_1=(\pi,0)$ and $\bm{Q}_2=(0,\pi)$ in the first line of Eq.~(\ref{eq:multiple-Q}). 
Here, $\bm{r}_i=(r_i^x,r_i^y)$, and we set the lattice constant as unity. 
Similarly, the triple-$Q$ state on a 3D cubic lattice in Fig.~\ref{Fig:ponti}(b) is given by $\bm{Q}_1=(0,\pi,\pi)$, $\bm{Q}_2=(\pi,0,\pi)$, and $\bm{Q}_3=(\pi,\pi,0)$ in the second line of Eq.~(\ref{eq:multiple-Q}). 
In the following, we focus on the multiple-$Q$ states with the shortest period, which have been studied in many previous studies~\cite{Yunoki:PhysRevB.62.13816,akagi2010spin,Alonso:PhysRevB.64.054408,Hayami:PhysRevB.89.085124}: $\bmod (Q_\eta^\mu,\pi) = 0$, where 
$\bm{Q}_\eta=(Q_\eta^x,Q_\eta^y,Q_\eta^z)$ ($\eta=1,2,3$). 
These states maximize the vector or scalar spin chirality, which leads to unusual transport properties. 
Table~\ref{tab} summarizes the lattice structure, multiple-$Q$ wave numbers, and corresponding rotational operation for the multiple-$Q$ orders considered in the present study.

\begin{table}[htb!]
\begin{tabular}{|c|c|c|}
\hline
 lattice  & multiple-$Q$ wave numbers& symmetry \\
 \hline
 \hline
 2D square & $(\pi,0)$, $(0,\pi)$ & $C_{4}$ \\
 \hline 
 2D triangular & $(\pi,0)$, $(0,\pi)$, $(\pi,\pi)$ & $C_{6}$ \\
 \hline
 3D cubic & $(0,\pi,\pi)$, $(\pi,0,\pi)$, $(\pi,\pi,0)$ & $C_{3} 
 $ \\
 \hline
 3D fcc & $(\pi,0,0)$, $(0,\pi,0)$, $(0,0,\pi)$ & $C_{3} 
 $ \\
 \hline
 \end{tabular}
 \caption{
Summary of the multiple-$Q$ magnetic orders discussed in this study. 
Schematic pictures of the spin patterns are shown in the insets of Fig.~\ref{Fig:FS}. 
 }
 \label{tab}
 \end{table}

Now we examine the instability of itinerant electrons toward the multiple-$Q$ orders given by Eq.~(\ref{eq:multiple-Q}). 
Specifically, we here consider noninteracting electrons coupled with localized spins by the local exchange coupling. 
The situation is described by the Kondo lattice Hamiltonian given by
\begin{align}
\label{eq:Ham}
\mathcal{H} = -t \sum_{\langle i,j \rangle \sigma} (c_{i\sigma}^{\dagger} c_{j \sigma} + {\rm H.c.}) - J \sum_{i \sigma \sigma'} c^{\dagger}_{i \sigma} \bm{\sigma}_{\sigma \sigma'} c_{i \sigma'} \cdot \bm{S}_i, 
\end{align}
where $c^{\dagger}_{i \sigma}$ ($c_{i \sigma}$) is a creation (annihilation) operator of itinerant electrons at site $i$ and spin $\sigma$, $\bm{\sigma} = (\sigma^x,\sigma^y,\sigma^z)$ is the vector of Pauli matrices, $\bm{S}_{i}$ is a localized spin at site $i$, and $J$ is the exchange coupling constant (the sign is irrelevant for the following arguments). 
The sum of $\langle i, j \rangle$ is taken over the nearest-neighbor sites. 
Hereafter, we take $t=1$ as an energy unit.
By replacing $\bm{S}_i$ by Eq.~(\ref{eq:multiple-Q}), the Hamiltonian is written in the form 
\begin{align}
\label{Ham_k}
\mathcal{H} = \sum_{\bm{k}\sigma} \epsilon_{\bm{k}} c_{\bm{k}\sigma}^{\dagger} c_{\bm{k}\sigma}
- J m 
\sum_{\bm{k}\sigma\sigma' \eta} c_{\bm{k} \sigma}^{\dagger} 
\sigma_{\sigma \sigma'}^\eta c_{\bm{k}+\bm{Q}_{\eta} \sigma'}, 
\end{align}
where $c_{\bm{k}\sigma}^{\dagger}$ and $c_{\bm{k}\sigma}$ are the 
Fourier transform of $c^{\dagger}_{i \sigma}$ and $c_{i \sigma}$, respectively; 
$\epsilon_{\bm{k}}$ is the energy dispersion for free electrons on each lattice. 
Here, $m$ is the normalization factor so that $|\langle \bm{S}_i \rangle|=1$: 
$m=1/\sqrt{2}$ and $1/\sqrt{3}$ for double- 
and triple-$Q$ states, respectively. 

\begin{figure}[hbt!]
\begin{center}
\includegraphics[width=1.0 \hsize]{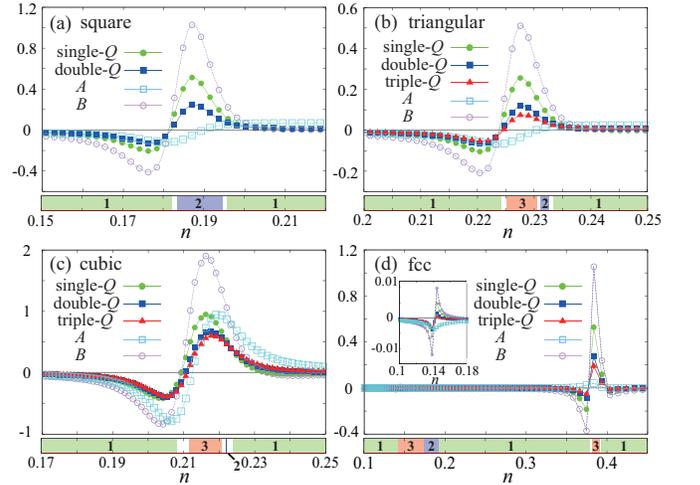} 
\caption{
\label{Fig:energy}
(Color online) 
Filling dependence of the fourth-order free energy $F^{(4)}_{\nu}$ in Eq.~(\ref{eq:F4multi}) divided by $J^4$ for the single-, double-, 
and triple-$Q$ magnetic ordered states: 
(a) square, (b) triangular, (c) cubic, and (d) fcc lattices. 
The parameters $A$ and $B$ are also plotted. 
The data are calculated at $T=0.03$. 
The inset of (d) shows the enlarged plot near $n=0.14$. 
The bottom strip in each figure represents the ground state phase diagram at $J=0.1$ obtained by the variational calculations; 
1, 2, 3 denote the single-, double-, 
and triple-$Q$ 
states, respectively, and  the white regions are the phase separation. 
}
\end{center}
\end{figure}

In order to clarify the dominant instability, we perform the perturbation expansion of the free energy in terms of $J$. 
The free energy $F$ is given by $F=-T\sum_{\nu} \log[1+\exp (-E_{\nu}/T)]$, where 
$E_{\nu}$ is the eigenvalues of $\mathcal{H}$, and $T$ is temperature 
(we set the Boltzmann constant $k_{\rm B}=1$). 
We consider the expansion up to the fourth order as 
$
F=F^{(0)}+F^{(2)}+F^{(4)} 
$. 
The second-order term $F^{(2)}$ is given in the general form, 
$
F^{(2)} = -J^2 \sum_{\bm{q}} |\bm{S}_{\bm{q}}|^2 \chi_0({\bm{q}}), 
$
where $\bm{S}_{\bm{q}}$ is the Fourier transform of $\bm{S}_i$, and $\chi_0({\bm{q}})$ is the bare susceptibility. 
$F^{(2)}$ has the form of the Ruderman-Kittel-Kasuya-Yosida interaction~\cite{Ruderman, Kasuya, Yosida1957}. 
The two-spin interaction induces the magnetic instability with wave vector $\bm{q}^*$ that maximizes $\chi_0 (\bm{q})$; 
$\bm{q}^*$ depends on the electron filling and, in general, is incommensurate to the lattice period.
On the other hand, $F^{(2)}$ for the single-, double-, and triple-$Q$ states are degenerate at all filling, 
which is written as 
\begin{align}
F^{(2)} = - 
(Jm)^2 T \sum_{\bm{k}\omega_p\eta} 
G^0_{\bm{k}+\bm{Q}_\eta} G^0_{\bm{k}}. 
\end{align}
Here, $G_{\bm{k}}^0 ({\rm i} \omega_{p})={[{\rm i} \omega_p -(\epsilon_{\bm{k}}-\mu)]}^{-1}$ is the noninteracting Green function, where $\omega_{p}$ is the Matsubara frequency and $\mu$ is the chemical potential; the single-$Q$ state is given by $\langle \bm{S}_i \rangle = (\cos \bm{Q}_{\eta}\cdot\bm{r}_i,0,0)$. 
At some particular filling, $\bm{q}^*$ coincides with $\bm{Q}_\eta$, i.e., the multiple-$Q$ states give the lowest energy within the second-order perturbation. 
At this filling, however, the degeneracy remains between the single-, double-, and triple-$Q$ states. 
In the following, we will discuss the possibility of multiple-$Q$ ordering in this degenerate situation. 

The degeneracy is lifted by the fourth-order perturbation. 
$F^{(4)}$ is given by
\begin{align}
\label{eq:F4}
F^{(4)} &= 
(Jm)^4 \frac{T}{2}  \sum_{\bm{k}\omega_p\eta\eta'} 
\!\! [G_{\bm{k}}^2
G_{\bm{k}+\bm{Q}_{\eta}}^2
\delta_{\eta \eta'} \nonumber 
 + (4 G_{\bm{k}}^2
G_{\bm{k}+\bm{Q}_{\eta}}
G_{\bm{k}+\bm{Q}_{\eta'}}
\nonumber \\
& \qquad - 2 G_{\bm{k}}
G_{\bm{k}+\bm{Q}_{\eta}}
G_{\bm{k}+\bm{Q}_{\eta'}}
G_{\bm{k}+\bm{Q}_{\eta}+\bm{Q}_{\eta'}} )
(1-\delta_{\eta \eta'}) ], 
\end{align}
which is specifically written as 
\begin{align}
\label{eq:F4multi}
F_{\nu}^{(4)} = \frac{J^4}{2} \left\{ \left(1-\frac{1}{\nu}\right)A + \frac{1}{\nu}B \right\}, 
\end{align}
for the single-$Q$ ($\nu=1$), double-$Q$ ($\nu=2$), and triple-$Q$ ($\nu=3$) states. 
In Eq.~(\ref{eq:F4multi}), 
\begin{align}
\label{eq:A}
A&=T \sum_{k}\sum_{\omega_{p}} 
(2G_{\bm{k}}^2
G_{\bm{k}+\bm{Q}_{\eta}}
G_{\bm{k}+\bm{Q}_{\eta'}}
\nonumber \\
&\qquad -G_{\bm{k}}
G_{\bm{k}+\bm{Q}_{\eta}}
G_{\bm{k}+\bm{Q}_{\eta'}}
G_{\bm{k}+\bm{Q}_{\eta}+\bm{Q}_{\eta'}}
), 
\\ 
B&=T \sum_{k}\sum_{\omega_{p}} G_{\bm{k}}^2
G_{\bm{k}+\bm{Q}}^2. 
\label{eq:B}
\end{align}
Here, $\eta \neq \eta'$ in Eq.~(\ref{eq:A}). 
These expressions are generic and independent of the lattice structures. 
Interestingly, the differences between any of two with different $\nu$ 
in Eq.~(\ref{eq:F4multi}) 
is proportional to $A-B$. 
This means that the phase transitions between the different states occur when $A=B$. 
The ground state is the triple(single)-$Q$ state for $A<(>)B$, as $F_{\rm triple}^{(4)} < F_{\rm double}^{(4)} < F_{\rm single}^{(4)}$ for $A<B$ and $F_{\rm triple}^{(4)} > F_{\rm double}^{(4)} > F_{\rm single}^{(4)}$ for $A>B$. 
In the 2D square case, however, the double-$Q$ state becomes the ground state for $A<B$, as 
the triple-$Q$ state is not included in the present consideration.

\begin{figure}[hbt!]
\begin{center}
\includegraphics[width=1.0 \hsize]{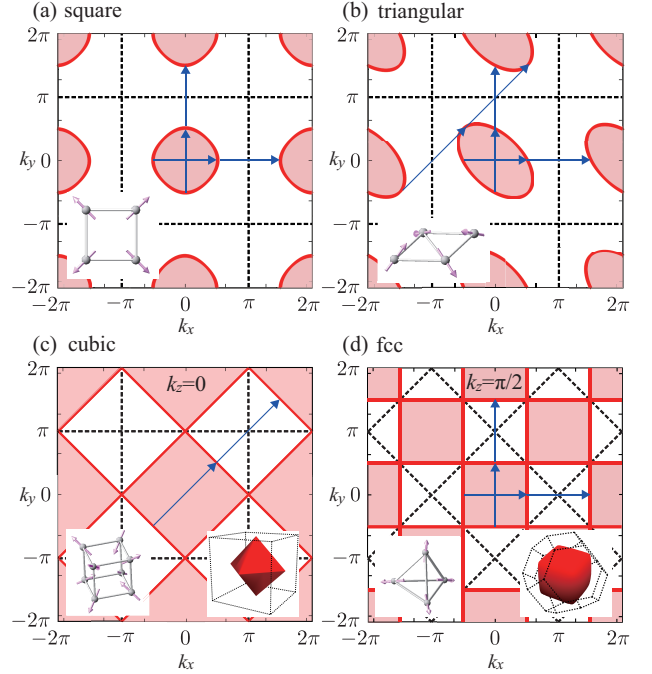} 
\caption{
\label{Fig:FS}
(Color online)
Fermi surfaces in the extended Brillouin zone scheme for (a) square ($\mu=-2$), (b) triangular ($\mu=-2$), (c) cubic ($\mu=-2$) at $k_z=0$, and (d) fcc lattices ($\mu=0$) at $k_z=\pi/2$. 
The arrows show the ($d-2$)-dimensional connections of the Fermi surfaces by the multiple-$Q$ wave vectors. 
The dashed lines show the Brillouin zone boundaries. 
In (c) and (d), the 3D pictures of the Fermi surfaces in the first Brillouin zone are presented. 
Each multiple-$Q$ magnetic ordered pattern is also shown in the inset of the figure. 
}
\end{center}
\end{figure}

Figure~\ref{Fig:energy} shows $F^{(4)} /J^4 $, $A$, and $B$ for the systems on 
square, triangular, cubic, and fcc lattices 
as functions of the electron filling $n =\frac{1}{2N}\sum_{i\sigma} \langle c_{i\sigma}^{\dagger}c_{i\sigma} \rangle$. 
The data are calculated at $T=0.03$. 
The results are qualitatively the same for different lattices; $F^{(4)} /J^4 
$ for the multiple-$Q$ state becomes lowest in a narrow window of $n$, while the single-$Q$ state is 
lower in energy for general filling. 
The narrow window in each case includes a special filling where $F^{(2)}$ becomes lowest for $\bm{q}^*=\bm{Q}_\eta$ and degenerate: 
$\mu^*=-2$ ($n\sim 0.185$) for square, $\mu^*=-2$ ($n\sim 0.226$) for triangular, 
$\mu^*=-2$ ($n\sim 0.213$) for cubic, and $\mu^*=-4$ ($n \sim 0.142$) and $0$ ($n \sim 0.383$) for fcc lattices. 
Note that there are two regions in the fcc case, and we focus on the stronger instability at $\mu^*=0$ in the following analysis~\cite{comment_fcc}. 
The results indicate that the multiple-$Q$ instabilities manifest in these situations where the second-order perturbation is degenerate. 

Such instabilities are indeed observed in the calculations beyond the perturbation theory. 
In Fig.~\ref{Fig:energy}, we compare the perturbation results with the ground states for small $J$ obtained by variational calculations. 
In the variational calculations, we determine the ground state numerically by comparing the free energy for the single-, double-, and triple-$Q$ ordered states 
(in the square lattice case, only single- and double-$Q$ states). 
The bottom strips in Fig.~\ref{Fig:energy} show the variational results at $J=0.1$. 
The results well agree with the perturbation results for $F^{(4)} /J^4 $; 
the multiple-$Q$ states appear in the variational ground states near the filling where the perturbation signals their instabilities.

We here discuss the reason why the multiple-$Q$ instabilities appear at the particular chemical potential $\mu^*$ by analyzing the electronic structure. 
Figure~\ref{Fig:FS} shows the FS at $\mu^*$ for each lattice in the extended Brillouin zone scheme. 
As shown in the figures, the FS at $\mu^*$ is connected by the multiple-$Q$ wave vectors not only within the Brillouin zone but also between the neighboring Brillouin zones simultaneously. 
Note that such connections are possible only for the wave vectors corresponding to the shortest period orders considered here. 
These connections are different from the perfect nesting, because the FS is connected partly, specifically, only a $(d-2)$-dimensional portion: zero-dimensional points in the 2D cases and one-dimensional lines in the 3D cases, as shown in Fig.~\ref{Fig:FS}~\cite{comment_fcc}. 
The $(d-2)$-dimensional connections of FS is the origin of the multiple-$Q$ instabilities. 

The multiple-$Q$ connections give rise to the critical divergence of the fourth-order contributions $F^{(4)} /J^4 
$ in Fig.~\ref{Fig:energy} in the limit of $T \rightarrow 0$. 
The divergence comes dominantly from $B$; 
we numerically find that $B \propto T^{-3/2 }$ 
for 2D and $B \propto T^{-5/3}$ for 3D. 
It leads to a local gap formation at the connected portions of the 
FS with the multiple-$Q$ ordering, which is relevant for stabilizing the multiple-$Q$ phases as found in the variational calculations. 
Although this was shown for the 2D triangular lattice case~\cite{akagi2012hidden}, we here find that the mechanism is universal, irrespective of the lattice structures as well as the system dimensions.

\begin{figure}[hbt!]
\begin{center}
\includegraphics[width=1.0 \hsize]{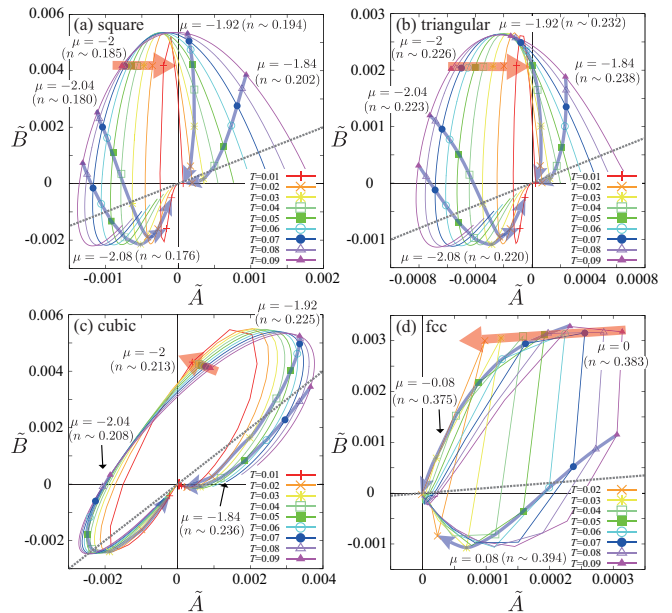} 
\caption{
\label{Fig:souzu}
(Color online)
The flow diagram of the two parameters $A$ and $B$ in Eqs.~(\ref{eq:A}) and (\ref{eq:B}), respectively, for (a) the square, (b) triangular, (c) cubic, and (d) fcc lattice cases. 
Each curve shows the trajectory of ($\tilde{A}, \tilde{B}$) while changing the chemical potential $\mu$ 
at a fixed $T$: 
$-2.4 \leq \mu < -1.6$ (every $0.005$) for (a) and (b), 
$-2.5 \leq \mu < -1.7$ (every $0.005$) for (c), and 
$-1.36 \leq \mu < 1.2$ (every $0.08$) for (d).
Here, $\tilde{A}=AT^{3/2}(AT^{5/3})$ and $\tilde{B}=BT^{3/2}(BT^{5/3})$ in (a) and (b) [(c) and (d)]. 
The arrows connecting the symbols indicate the typical flows as $T \to 0$ at the fixed $\mu$ whose values are indicated in the figures. 
The dashed straight lines $\tilde{A}=\tilde{B}$ show the phase boundaries, above which the multiple-$Q$ instabilities occur. 
Hence, the thick arrows approaching a nonzero $\tilde{B}$ signal multiple-$Q$ instabilities at 
$\mu=\mu^*$, while the other flows for $\mu \neq \mu^*$ merge to the origin, indicating that the single-$Q$ order is more stable. 
}
\end{center}
\end{figure}

We summarize the multiple-$Q$ instabilities in the form of the flow diagram on the plane of $A$ and $B$. 
Figure~\ref{Fig:souzu} shows the trajectories of the renomarlized parameters $\tilde{A}$ and $\tilde{B}$ while changing 
$\mu$ at fixed $T$; here, $\tilde{A}=AT^{3/2}(AT^{5/3})$ and $\tilde{B}=BT^{3/2}(BT^{5/3})$ for 2D (3D). 
When we trace the data for a fixed $\mu$, all the flows merge to the origin $(\tilde{A},\tilde{B})=(0,0)$, except for the flow corresponding to the particular $\mu^*$. 
In other words, the flow approaching a ``fixed point" with a nonzero value of $\tilde{B}$ signals the multiple-$Q$ instability. 
This flow diagram provides a universal tool to detect the multiple-$Q$ instabilities of itinerant electrons coupled to localized spins.

To summarize, we have investigated the stabilization mechanism of multiple-$Q$ orders showing noncollinear and noncoplanar spin textures in itinerant magnets. 
By carefully analyzing the fourth-order perturbation in terms of the spin-charge coupling, we have found that the multiple-$Q$ instabilities ubiquitously appear at particular filling in various lattice structures. 
We have found that the instabilities are caused by ($d-2$)-dimensional connections of the Fermi surfaces in the extended Brillouin zone scheme. 
The results are graphically summarized in a flow diagram for the two parameters appearing in the fourth-order perturbations. 
Our results provide a new robust mechanism of noncollinear and noncoplanar orders originating from the simple Fermi surface topology. 
The mechanism might be extended to multi-orbital systems, possibly leading to multiple-$Q$ spin-orbital ordering with a longer period. 
An extension to non-Bravais lattices is also interesting. 
Such extensions of the ($d-2$)-dimensional connection scenario will stimulate further exploration of exotic phenomena in multiple-$Q$ ordering. 

The authors acknowledge Yutaka Akagi for fruitful discussions. 
SH is supported by Grant-in-Aid for JSPS Fellows. 
This work was supported by Grants-in-Aid for Scientific Research (No.~24340076), the Strategic Programs for Innovative Research (SPIRE), MEXT, and the Computational Materials Science Initiative (CMSI), Japan.

\bibliographystyle{apsrev}
\bibliography{ref}

\begin{thebibliography}{30}
\expandafter\ifx\csname natexlab\endcsname\relax\def\natexlab#1{#1}\fi
\expandafter\ifx\csname bibnamefont\endcsname\relax
  \def\bibnamefont#1{#1}\fi
\expandafter\ifx\csname bibfnamefont\endcsname\relax
  \def\bibfnamefont#1{#1}\fi
\expandafter\ifx\csname citenamefont\endcsname\relax
  \def\citenamefont#1{#1}\fi
\expandafter\ifx\csname url\endcsname\relax
  \def\url#1{\texttt{#1}}\fi
\expandafter\ifx\csname urlprefix\endcsname\relax\def\urlprefix{URL }\fi
\providecommand{\bibinfo}[2]{#2}
\providecommand{\eprint}[2][]{\url{#2}}

\bibitem[{\citenamefont{Peierls}(1955)}]{peierls1955quantum}
\bibinfo{author}{\bibfnamefont{R.~E.} \bibnamefont{Peierls}},
  \emph{\bibinfo{title}{Quantum theory of solids}}, \bibinfo{number}{23}
  (\bibinfo{publisher}{Oxford University Press}, \bibinfo{year}{1955}).

\bibitem[{\citenamefont{Gr\"uner}(1988)}]{RevModPhys.60.1129}
\bibinfo{author}{\bibfnamefont{G.}~\bibnamefont{Gr\"uner}},
  \bibinfo{journal}{Rev. Mod. Phys.} \textbf{\bibinfo{volume}{60}},
  \bibinfo{pages}{1129} (\bibinfo{year}{1988}).

\bibitem[{\citenamefont{Gr\"uner}(1994)}]{RevModPhys.66.1}
\bibinfo{author}{\bibfnamefont{G.}~\bibnamefont{Gr\"uner}},
  \bibinfo{journal}{Rev. Mod. Phys.} \textbf{\bibinfo{volume}{66}},
  \bibinfo{pages}{1} (\bibinfo{year}{1994}).

\bibitem[{\citenamefont{Martin and
  Batista}(2008)}]{Martin:PhysRevLett.101.156402}
\bibinfo{author}{\bibfnamefont{I.}~\bibnamefont{Martin}} \bibnamefont{and}
  \bibinfo{author}{\bibfnamefont{C.~D.} \bibnamefont{Batista}},
  \bibinfo{journal}{Phys. Rev. Lett.} \textbf{\bibinfo{volume}{101}},
  \bibinfo{pages}{156402} (\bibinfo{year}{2008}).

\bibitem[{\citenamefont{Chern}(2010)}]{Chern:PhysRevLett.105.226403}
\bibinfo{author}{\bibfnamefont{G.-W.} \bibnamefont{Chern}},
  \bibinfo{journal}{Phys. Rev. Lett.} \textbf{\bibinfo{volume}{105}},
  \bibinfo{pages}{226403} (\bibinfo{year}{2010}).

\bibitem[{\citenamefont{Venderbos et~al.}(2012)\citenamefont{Venderbos,
  Daghofer, van~den Brink, and Kumar}}]{Venderbos:PhysRevLett.109.166405}
\bibinfo{author}{\bibfnamefont{J.~W.~F.} \bibnamefont{Venderbos}},
  \bibinfo{author}{\bibfnamefont{M.}~\bibnamefont{Daghofer}},
  \bibinfo{author}{\bibfnamefont{J.}~\bibnamefont{van~den Brink}},
  \bibnamefont{and} \bibinfo{author}{\bibfnamefont{S.}~\bibnamefont{Kumar}},
  \bibinfo{journal}{Phys. Rev. Lett.} \textbf{\bibinfo{volume}{109}},
  \bibinfo{pages}{166405} (\bibinfo{year}{2012}).

\bibitem[{\citenamefont{Loss and Goldbart}(1992)}]{Loss_PhysRevB.45.13544}
\bibinfo{author}{\bibfnamefont{D.}~\bibnamefont{Loss}} \bibnamefont{and}
  \bibinfo{author}{\bibfnamefont{P.~M.} \bibnamefont{Goldbart}},
  \bibinfo{journal}{Phys. Rev. B} \textbf{\bibinfo{volume}{45}},
  \bibinfo{pages}{13544} (\bibinfo{year}{1992}).

\bibitem[{\citenamefont{Ye et~al.}(1999)\citenamefont{Ye, Kim, Millis,
  Shraiman, Majumdar, and Te\ifmmode \check{s}\else
  \v{s}\fi{}anovi\ifmmode~\acute{c}\else \'{c}\fi{}}}]{Ye_PhysRevLett.83.3737}
\bibinfo{author}{\bibfnamefont{J.}~\bibnamefont{Ye}},
  \bibinfo{author}{\bibfnamefont{Y.~B.} \bibnamefont{Kim}},
  \bibinfo{author}{\bibfnamefont{A.~J.} \bibnamefont{Millis}},
  \bibinfo{author}{\bibfnamefont{B.~I.} \bibnamefont{Shraiman}},
  \bibinfo{author}{\bibfnamefont{P.}~\bibnamefont{Majumdar}}, \bibnamefont{and}
  \bibinfo{author}{\bibfnamefont{Z.}~\bibnamefont{Te\ifmmode \check{s}\else
  \v{s}\fi{}anovi\ifmmode~\acute{c}\else \'{c}\fi{}}}, \bibinfo{journal}{Phys.
  Rev. Lett.} \textbf{\bibinfo{volume}{83}}, \bibinfo{pages}{3737}
  (\bibinfo{year}{1999}).

\bibitem[{\citenamefont{Ohgushi et~al.}(2000)\citenamefont{Ohgushi, Murakami,
  and Nagaosa}}]{Ohgushi:PhysRevB.62.R6065}
\bibinfo{author}{\bibfnamefont{K.}~\bibnamefont{Ohgushi}},
  \bibinfo{author}{\bibfnamefont{S.}~\bibnamefont{Murakami}}, \bibnamefont{and}
  \bibinfo{author}{\bibfnamefont{N.}~\bibnamefont{Nagaosa}},
  \bibinfo{journal}{Phys. Rev. B} \textbf{\bibinfo{volume}{62}},
  \bibinfo{pages}{R6065} (\bibinfo{year}{2000}).

\bibitem[{\citenamefont{Taguchi et~al.}(2001)\citenamefont{Taguchi, Oohara,
  Yoshizawa, Nagaosa, and Tokura}}]{taguchi2001spin}
\bibinfo{author}{\bibfnamefont{Y.}~\bibnamefont{Taguchi}},
  \bibinfo{author}{\bibfnamefont{Y.}~\bibnamefont{Oohara}},
  \bibinfo{author}{\bibfnamefont{H.}~\bibnamefont{Yoshizawa}},
  \bibinfo{author}{\bibfnamefont{N.}~\bibnamefont{Nagaosa}}, \bibnamefont{and}
  \bibinfo{author}{\bibfnamefont{Y.}~\bibnamefont{Tokura}},
  \bibinfo{journal}{Science} \textbf{\bibinfo{volume}{291}},
  \bibinfo{pages}{2573} (\bibinfo{year}{2001}).

\bibitem[{\citenamefont{Taguchi and Tatara}(2009)}]{TaguchiPhysRevB.79.054423}
\bibinfo{author}{\bibfnamefont{K.}~\bibnamefont{Taguchi}} \bibnamefont{and}
  \bibinfo{author}{\bibfnamefont{G.}~\bibnamefont{Tatara}},
  \bibinfo{journal}{Phys. Rev. B} \textbf{\bibinfo{volume}{79}},
  \bibinfo{pages}{054423} (\bibinfo{year}{2009}).

\bibitem[{\citenamefont{Agterberg and Yunoki}(2000)}]{Yunoki:PhysRevB.62.13816}
\bibinfo{author}{\bibfnamefont{D.~F.} \bibnamefont{Agterberg}}
  \bibnamefont{and} \bibinfo{author}{\bibfnamefont{S.}~\bibnamefont{Yunoki}},
  \bibinfo{journal}{Phys. Rev. B} \textbf{\bibinfo{volume}{62}},
  \bibinfo{pages}{13816} (\bibinfo{year}{2000}).

\bibitem[{\citenamefont{Alonso et~al.}(2001)\citenamefont{Alonso, Capit\'an,
  Fern\'andez, Guinea, and Mart\'in-Mayor}}]{Alonso:PhysRevB.64.054408}
\bibinfo{author}{\bibfnamefont{J.~L.} \bibnamefont{Alonso}},
  \bibinfo{author}{\bibfnamefont{J.~A.} \bibnamefont{Capit\'an}},
  \bibinfo{author}{\bibfnamefont{L.~A.} \bibnamefont{Fern\'andez}},
  \bibinfo{author}{\bibfnamefont{F.}~\bibnamefont{Guinea}}, \bibnamefont{and}
  \bibinfo{author}{\bibfnamefont{V.}~\bibnamefont{Mart\'in-Mayor}},
  \bibinfo{journal}{Phys. Rev. B} \textbf{\bibinfo{volume}{64}},
  \bibinfo{pages}{054408} (\bibinfo{year}{2001}).

\bibitem[{\citenamefont{Hayami et~al.}(2014)\citenamefont{Hayami, Misawa,
  Yamaji, and Motome}}]{Hayami:PhysRevB.89.085124}
\bibinfo{author}{\bibfnamefont{S.}~\bibnamefont{Hayami}},
  \bibinfo{author}{\bibfnamefont{T.}~\bibnamefont{Misawa}},
  \bibinfo{author}{\bibfnamefont{Y.}~\bibnamefont{Yamaji}}, \bibnamefont{and}
  \bibinfo{author}{\bibfnamefont{Y.}~\bibnamefont{Motome}},
  \bibinfo{journal}{Phys. Rev. B} \textbf{\bibinfo{volume}{89}},
  \bibinfo{pages}{085124} (\bibinfo{year}{2014}).

\bibitem[{\citenamefont{Akagi and Motome}(2010)}]{akagi2010spin}
\bibinfo{author}{\bibfnamefont{Y.}~\bibnamefont{Akagi}} \bibnamefont{and}
  \bibinfo{author}{\bibfnamefont{Y.}~\bibnamefont{Motome}},
  \bibinfo{journal}{J. Phys. Soc. Jpn.} \textbf{\bibinfo{volume}{79}}
  (\bibinfo{year}{2010}).

\bibitem[{\citenamefont{Kumar and van~den
  Brink}(2010)}]{Kumar:PhysRevLett.105.216405}
\bibinfo{author}{\bibfnamefont{S.}~\bibnamefont{Kumar}} \bibnamefont{and}
  \bibinfo{author}{\bibfnamefont{J.}~\bibnamefont{van~den Brink}},
  \bibinfo{journal}{Phys. Rev. Lett.} \textbf{\bibinfo{volume}{105}},
  \bibinfo{pages}{216405} (\bibinfo{year}{2010}).

\bibitem[{\citenamefont{Kato et~al.}(2010)\citenamefont{Kato, Martin, and
  Batista}}]{Kato:PhysRevLett.105.266405}
\bibinfo{author}{\bibfnamefont{Y.}~\bibnamefont{Kato}},
  \bibinfo{author}{\bibfnamefont{I.}~\bibnamefont{Martin}}, \bibnamefont{and}
  \bibinfo{author}{\bibfnamefont{C.~D.} \bibnamefont{Batista}},
  \bibinfo{journal}{Phys. Rev. Lett.} \textbf{\bibinfo{volume}{105}},
  \bibinfo{pages}{266405} (\bibinfo{year}{2010}).

\bibitem[{\citenamefont{Fiebig}(2005)}]{Fiebig:0022-3727-38-8-R01}
\bibinfo{author}{\bibfnamefont{M.}~\bibnamefont{Fiebig}}, \bibinfo{journal}{J.
  Phys. D: Appl. Phys.} \textbf{\bibinfo{volume}{38}}, \bibinfo{pages}{R123}
  (\bibinfo{year}{2005}).

\bibitem[{\citenamefont{Katsura et~al.}(2005)\citenamefont{Katsura, Nagaosa,
  and Balatsky}}]{Katsura:PhysRevLett.95.057205}
\bibinfo{author}{\bibfnamefont{H.}~\bibnamefont{Katsura}},
  \bibinfo{author}{\bibfnamefont{N.}~\bibnamefont{Nagaosa}}, \bibnamefont{and}
  \bibinfo{author}{\bibfnamefont{A.~V.} \bibnamefont{Balatsky}},
  \bibinfo{journal}{Phys. Rev. Lett.} \textbf{\bibinfo{volume}{95}},
  \bibinfo{pages}{057205} (\bibinfo{year}{2005}).

\bibitem[{\citenamefont{Sergienko and
  Dagotto}(2006)}]{Sergienko_PhysRevB.73.094434}
\bibinfo{author}{\bibfnamefont{I.~A.} \bibnamefont{Sergienko}}
  \bibnamefont{and} \bibinfo{author}{\bibfnamefont{E.}~\bibnamefont{Dagotto}},
  \bibinfo{journal}{Phys. Rev. B} \textbf{\bibinfo{volume}{73}},
  \bibinfo{pages}{094434} (\bibinfo{year}{2006}).

\bibitem[{\citenamefont{Jia et~al.}(2007)\citenamefont{Jia, Onoda, Nagaosa, and
  Han}}]{Jia:PhysRevB.76.144424}
\bibinfo{author}{\bibfnamefont{C.}~\bibnamefont{Jia}},
  \bibinfo{author}{\bibfnamefont{S.}~\bibnamefont{Onoda}},
  \bibinfo{author}{\bibfnamefont{N.}~\bibnamefont{Nagaosa}}, \bibnamefont{and}
  \bibinfo{author}{\bibfnamefont{J.~H.} \bibnamefont{Han}},
  \bibinfo{journal}{Phys. Rev. B} \textbf{\bibinfo{volume}{76}},
  \bibinfo{pages}{144424} (\bibinfo{year}{2007}).

\bibitem[{\citenamefont{Khomskii}(2009)}]{khomskii2009classifying}
\bibinfo{author}{\bibfnamefont{D.}~\bibnamefont{Khomskii}},
  \bibinfo{journal}{Physics} \textbf{\bibinfo{volume}{2}}, \bibinfo{pages}{20}
  (\bibinfo{year}{2009}).

\bibitem[{\citenamefont{Mostovoy}(2006)}]{PhysRevLett.96.067601}
\bibinfo{author}{\bibfnamefont{M.}~\bibnamefont{Mostovoy}},
  \bibinfo{journal}{Phys. Rev. Lett.} \textbf{\bibinfo{volume}{96}},
  \bibinfo{pages}{067601} (\bibinfo{year}{2006}).

\bibitem[{\citenamefont{M{\"u}hlbauer et~al.}(2009)\citenamefont{M{\"u}hlbauer,
  Binz, Jonietz, Pfleiderer, Rosch, Neubauer, Georgii, and
  B{\"o}ni}}]{muhlbauer2009skyrmion}
\bibinfo{author}{\bibfnamefont{S.}~\bibnamefont{M{\"u}hlbauer}},
  \bibinfo{author}{\bibfnamefont{B.}~\bibnamefont{Binz}},
  \bibinfo{author}{\bibfnamefont{F.}~\bibnamefont{Jonietz}},
  \bibinfo{author}{\bibfnamefont{C.}~\bibnamefont{Pfleiderer}},
  \bibinfo{author}{\bibfnamefont{A.}~\bibnamefont{Rosch}},
  \bibinfo{author}{\bibfnamefont{A.}~\bibnamefont{Neubauer}},
  \bibinfo{author}{\bibfnamefont{R.}~\bibnamefont{Georgii}}, \bibnamefont{and}
  \bibinfo{author}{\bibfnamefont{P.}~\bibnamefont{B{\"o}ni}},
  \bibinfo{journal}{Science} \textbf{\bibinfo{volume}{323}},
  \bibinfo{pages}{915} (\bibinfo{year}{2009}).

\bibitem[{\citenamefont{Yu et~al.}(2010)\citenamefont{Yu, Onose, Kanazawa,
  Park, Han, Matsui, Nagaosa, and Tokura}}]{yu2010real}
\bibinfo{author}{\bibfnamefont{X.}~\bibnamefont{Yu}},
  \bibinfo{author}{\bibfnamefont{Y.}~\bibnamefont{Onose}},
  \bibinfo{author}{\bibfnamefont{N.}~\bibnamefont{Kanazawa}},
  \bibinfo{author}{\bibfnamefont{J.}~\bibnamefont{Park}},
  \bibinfo{author}{\bibfnamefont{J.}~\bibnamefont{Han}},
  \bibinfo{author}{\bibfnamefont{Y.}~\bibnamefont{Matsui}},
  \bibinfo{author}{\bibfnamefont{N.}~\bibnamefont{Nagaosa}}, \bibnamefont{and}
  \bibinfo{author}{\bibfnamefont{Y.}~\bibnamefont{Tokura}},
  \bibinfo{journal}{Nature} \textbf{\bibinfo{volume}{465}},
  \bibinfo{pages}{901} (\bibinfo{year}{2010}).

\bibitem[{\citenamefont{Akagi et~al.}(2012)\citenamefont{Akagi, Udagawa, and
  Motome}}]{akagi2012hidden}
\bibinfo{author}{\bibfnamefont{Y.}~\bibnamefont{Akagi}},
  \bibinfo{author}{\bibfnamefont{M.}~\bibnamefont{Udagawa}}, \bibnamefont{and}
  \bibinfo{author}{\bibfnamefont{Y.}~\bibnamefont{Motome}},
  \bibinfo{journal}{Phys. Rev. Lett.} \textbf{\bibinfo{volume}{108}},
  \bibinfo{pages}{096401} (\bibinfo{year}{2012}).

\bibitem[{\citenamefont{Ruderman and Kittel}(1954)}]{Ruderman}
\bibinfo{author}{\bibfnamefont{M.~A.} \bibnamefont{Ruderman}} \bibnamefont{and}
  \bibinfo{author}{\bibfnamefont{C.}~\bibnamefont{Kittel}},
  \bibinfo{journal}{Phys. Rev.} \textbf{\bibinfo{volume}{96}},
  \bibinfo{pages}{99} (\bibinfo{year}{1954}).

\bibitem[{\citenamefont{Kasuya}(1956)}]{Kasuya}
\bibinfo{author}{\bibfnamefont{T.}~\bibnamefont{Kasuya}},
  \bibinfo{journal}{Prog. Theor. Phys.} \textbf{\bibinfo{volume}{16}},
  \bibinfo{pages}{45} (\bibinfo{year}{1956}).

\bibitem[{\citenamefont{Yosida}(1957)}]{Yosida1957}
\bibinfo{author}{\bibfnamefont{K.}~\bibnamefont{Yosida}},
  \bibinfo{journal}{Phys. Rev.} \textbf{\bibinfo{volume}{106}},
  \bibinfo{pages}{893} (\bibinfo{year}{1957}).

\bibitem[{com()}]{comment_fcc}
\bibinfo{note}{The weak anomaly at $\mu^*=-4$ ($n\sim 0.142$) for fcc in
  Fig.~\ref{Fig:energy}(d) originates in the point connections of FS. A similar
  weak anomaly is seen also for cubic at $\mu^*=-4$ ($n\sim 0.057$) with
  $\bm{Q}_1 =(\pi,0,0)$, $\bm{Q}_2 =(0,\pi,0)$, $\bm{Q}_3 =(0,0,\pi)$.}

\end{thebibliography}

\end{document}